\documentstyle[12pt]{article}
\makeatletter
\@addtoreset{equation}{section}
\makeatother

\begin{document}
\begin{center}
{\Large \bf
Indeterministic Quantum Gravity and Cosmology} \\[0.5cm]
{\large\bf IX. Nonreality of Many-Place Gravitational
Autolocalization: Why a Ball Is Not Located in Different
Places at Once}\\[1.5cm]
{\bf Vladimir S.~MASHKEVICH}\footnote {E-mail:
mash@gluk.apc.org}  \\[1.4cm]
{\it Institute of Physics, National academy
of sciences of Ukraine \\
252028 Kiev, Ukraine} \\[1.4cm]
\vskip 1cm

{\large \bf Abstract}
\end{center}

This paper is a sequel of papers [1-8], being an
immediate continuation and supplement to the last of
them, where gravitational autolocalization of a body has
been considered. A resulting solution, which describes a
one-place location, has been called gravilon. Here it is
shown that a gravilon is the only solution, i.e., that
many-place gravitational autolocalization is unreal.
This is closely related to nonreality of tunneling
in the conditions under consideration.

\newpage

\hspace*{5 cm}
\begin{minipage} [b] {12 cm}
Why, then, do we not experience macroscopic bodies, say
cricket balls, or even people, having two completely different
locations at once? This is a profound question, and present-day
quantum theory does not really provide us with a satisfactory
answer.
\end{minipage}
\begin{flushright}
Roger Penrose \vspace*{0.8 cm}
\end{flushright}

\begin{flushleft}
\hspace*{0.5 cm} {\Large \bf Introduction}
\end{flushleft}

In the previous paper [8] of this series, the possibility of
a gravitational autolocalization has been established. A
resulting solution describes a one-place location of a body,
which is called gravilon. But in [8] the question on nonreality
of many-place gravitational autolocalization has not been
even raised. This paper makes up for this essential
deficiency.

The answer to the question is positive: Many-place
gravitational autolocalization is unreal. The essence of the
matter is as follows.

For the sake of simplicity, we consider the case of a two-place
location. A wave function of the center of mass of a ball in
such a situation depends on some parameters: the distance
between location places and coefficients in the linear
combination of one-place functions. The requirement that the
wave function be an eigenfunction of the Hamiltonian results
in a relation for the parameters, which singles out a set of
measure zero for the admissible values of the parameters.
This result is obtained neglecting the tunneling between
the locations, so that nonreality of many-place location is
closely related to nonreality of tunneling in the
conditions under consideration.

Thus the only solution to the problem of gravitational
autolocalization is a one-place location, i.e., a gravilon.

\section{Basic equation}

Let $\psi(\vec r)$ be a wave function of the center of mass
of a ball with a mass $M$ [8]. $\psi$ must satisfy the
Schr$\ddot {\rm o}$dinger equation,
\begin{equation}
H\psi=\epsilon\psi
\label{1.1}
\end{equation}
where the Hamiltonian
\begin{equation}
H\equiv H[\psi]=-\frac{\hbar^{2}}{2M}\triangle+V(\vec r;\psi)
\label{1.2}
\end{equation}
and $V$ is the potential energy of the ball in a gravitational
field caused by the ball itself with the wave function $\psi$.

\section{Two-place function}

For the sake of simplicity, let us consider the case of a
two-place location:
\begin{equation}
\psi\equiv
\psi(\vec r)=c_{\rm I}\psi_{\rm I}(\vec r)+
c_{\rm II}\psi_{\rm II}(\vec r)
\label{2.1}
\end{equation}
where
\begin{equation}
\psi_{\rm I}=\psi_{1}(\vec r),\quad \psi_{\rm II}=
\psi_{2}(\vec r-\vec R)
\label{2.2}
\end{equation}
and
\begin{equation}
R\gg a_{0},\:r_{0i},\qquad i=1,2,
\label{2.3}
\end{equation}
here $a_{0}$ is the radius of the ball and $r_{0i}$ is an
effective radius of $\psi_{i}$. We may put
\begin{equation}
c_{\rm I}=e^{{\rm i}\alpha}\sqrt{1-b^{2}},\qquad c_{\rm II}=
e^{{\rm i}\alpha}e^{{\rm i}\beta}b,\qquad 0\le b\le 1,
\label{2.4}
\end{equation}
so that, dropping $e^{{\rm i}\alpha}$, we have
\begin{equation}
\psi\equiv\psi(\vec r;b,\beta,\vec R)=
\sqrt{1-b^{2}}\psi_{1}(\vec r)+e^{{\rm i}\beta}b
\psi_{2}(\vec r-\vec R).
\label{2.5}
\end{equation}

Now, in view of eq.(\ref{2.3}), we neglect the tunneling
between the two locations and choose $\psi_{1},\:\psi_{2}$
to be eigenfunctions of the Hamiltonian,
\begin{equation}
H\psi_{i}=\epsilon_{i}\psi_{i}.
\label{2.6}
\end{equation}

We have
\begin{equation}
H=H[b,\beta,\vec R],\;\psi_{1}=\psi_{1}(\vec r;b,\beta,\vec R),
\;\psi_{2}=\psi_{2}(\vec r-\vec R;b,\beta,\vec R),
\;\epsilon_{i}=\epsilon_{i}(b,\beta,\vec R).
\label{2.7}
\end{equation}

\section{The set of two-place states: measure zero}

We have from eqs.(\ref{2.5}), (\ref{2.6})
\begin{equation}
H\psi=\sqrt{1-b^{2}}\epsilon_{1}\psi_{1}+
e^{{\rm i}\beta}b\epsilon_{2}\psi_{2},
\label{3.1}
\end{equation}
so that for (\ref{1.1}) to be satisfied, the relation
\begin{equation}
\epsilon_{1}(b,\beta,\vec R)=\epsilon_{2}(b,\beta,\vec R)
\equiv \epsilon(b,\beta,\vec R)
\label{3.2}
\end{equation}
must be fulfilled.

Eq.(\ref{3.2}) determines a set of measure zero for
admissible values of the parameters $b,\beta$, and $\vec R$.
This implies that many-place gravitational autolocalization
is unreal.

\section{Tunneling}

Let us consider the tunneling, which was neglected. The
condition of this neglect is
\begin{equation}
DV\ll \Delta\epsilon
\label{4.1}
\end{equation}
where
\begin{equation}
D\sim e^{-2\sqrt{2M(V-\epsilon)/\hbar^{2}}L},
\label{4.2}
\end{equation}
is the transmission coefficient and $\Delta\epsilon$ is the
distance between energy levels.

We have [8]
\begin{equation}
V-\epsilon\approx V\sim \frac{\kappa M^{2}}{a},\qquad a=
a_{0}+r_{0},
\label{4.3}
\end{equation}
where $\kappa$ is the gravitational constant,
\begin{equation}
\Delta\epsilon\sim \frac{\hbar^{2}}{Ma^{2}},
\label{4.4}
\end{equation}
and
\begin{equation}
L=R.
\label{4.5}
\end{equation}
Thus eq.(\ref{4.1}) reduces to
\begin{equation}
\lambda e^{-\sqrt{\lambda}R/a}\ll 1
\label{4.6}
\end{equation}
with
\begin{equation}
\lambda\sim \frac{\kappa M^{3}}{\hbar^{2}}a,
\label{4.7}
\end{equation}
or
\begin{equation}
\lambda\sim\left( \frac{M}{m_{P}} \right)^{3}
\frac{a}{l_{P}},
\label{4.8}
\end{equation}
where $m_{P}$ is the Planck mass and $l_{P}$ is the
Planck length.

We have by eq.(\ref{2.3})
\begin{equation}
\frac{R}{a}\gg 1,
\label{4.9}
\end{equation}
so that the condition (\ref{4.6}) is fulfilled. In fact,
since by [8]
\begin{equation}
\lambda\gg 1
\label{4.10}
\end{equation}
holds, the inequality (\ref{4.6}) is very strong.

Thus nonreality of many-place gravitational autolocalization
is closely related to nonreality of tunneling in the
conditions under consideration.

\section*{Acknowledgment}

I would like to thank Stefan V. Mashkevich for helpful
discussions.

\end{document}